\documentclass[9pt]{article}
\usepackage[utf8]{inputenc}
\usepackage{graphicx}
\usepackage{subcaption}
\usepackage{amsthm}
\usepackage{mathrsfs}
\usepackage{slashed}
\usepackage{braket}
\usepackage{amsmath}
\begin{document}
\title{\bf Primordial thermal gravitational waves from warm inflation}
\date{}
\maketitle
\begin{center}\vspace{-15mm}
Anupama B\footnote{Corresponding author : 21phph19@uohyd.ac.in}\\
\small{School of Physics, University of Hyderabad, \\ P.O. Central University, Hyderabad  - 500046, India.\\ $^1$Corresponding author : 21phph19@uohyd.ac.in
}
\end{center}\vspace{10mm}
\begin{abstract}
For the first time, the possibility of generation of primordial thermal gravitational waves from warm inflation is investigated with the cosmic microwave background. The combined effect of thermal and squeezing effects can reduce the amplitude of the CMB BB-mode angular power spectrum for the parameter range ($0 \leq q \leq 1$ and $\zeta = 0,\pi/2$) considered in this work in both strong ($Q>>1$) and weak ($Q<<1$) warm inflationary models. In a thermal squeezed vacuum state, the BB-mode correlation angular power spectrum of warm inflation is found to be consistent with the dust-corrected joint constraints from $Planck$ and BK18 for both strong and weak dissipation regimes. The combined effect of thermal and squeezing phenomena in warm inflation is essential for achieving compatibility with current CMB BB-mode observations and predicts primordial thermal gravitational waves.
\end{abstract}
\vspace{1pc}
\section{Introduction}
Inflation was proposed to explain the observed large scale homogeneity, isotropy, and near flatness of the universe, which are difficult to account for within the standard Big Bang framework alone. Although the canonical cold inflation (CI) successfully explains some of the major problems of the standard model of cosmology, it is still plagued by issues related to the mechanism of reheating and the transition of the universe from quantum to classical. Most of the conventional models of inflation are ruled out when compared with the improved cosmic microwave background (CMB) observations. Therefore, recent studies focus on fabricating models that can fit the observations. This resulted in the development of a plethora of  CI models \cite{enc}. Although many cold inflation models are phenomenologically consistent with current observations \cite{planck}, constructing theoretically well motivated scenario to explain their microphysical realizations often requires significant fine tuning \cite{DB, eft} of the inflaton potential, careful control of radiative corrections \cite{sta}, or super-Planckian field excursions. A major theoretical challenge in constructing viable inflationary models is the $\eta$-problem \cite{eta1, eta2} , which arises from the difficulty of maintaining the flatness of the inflaton potential against quantum and gravitational corrections.  In a generic effective field theory, Planck-suppressed operators and radiative corrections typically induce corrections of the order of Hubble scale to the inflaton mass. Consequently, the inflaton potential becomes too steep to sustain prolonged slow-roll inflation without invoking additional symmetries or fine tuning. Hence, they are incompatible with swampland conjectures \cite{swa1, swa2, swa3}. This poses a difficulty in testing quantum gravity through inflation. 
Although primordial gravitational waves have not yet been directly detected, this absence is primarily associated with the exceptionally weak amplitude of the primordial tensor signal and the formidable observational challenge of distinguishing it from instrumental and astrophysical backgrounds. Their potential signatures can nevertheless be constrained indirectly through observables such as the B-mode polarization of the CMB.
Previous studies on the existence of gravitons in thermal squeezed states in CI lack a solid foundation to explain the generation of thermal effect, nor did they shed any light on the B modes, as they were studied using WMAP and other data \cite{prek1,prek2}. This motivates to explore a more comprehensive, physically grounded framework with a consistent description of inflationary microphysics and later stages, such as reheating, in the light of recent observational data from the joint analysis of $Planck$ and BICEP2/$Keck$. Warm inflation (WI) is such an alternate scenario which assumes a non negligible radiation at the beginning of inflation and the subsequent interaction of the inflaton with other fields prevents supercooling in the early universe \cite{review, ori1, ori2, ori3, ori4}. In addition to the reheating problem, WI is also found to play a role in alleviating one of the major tensions in cosmology known as the Hubble tension and is now gaining popularity by accommodating the models that were once ruled out by CI \cite{cqg2}.  It is worth exploring the four major models of WI, the Distributed Mass Model (DMM) originating from string theory \cite{dmmori1,dmmori2}, Two Stage Inflation (TSI) involving supersymmetry \cite{tsiori1,tsiori2}, Warm Little Inflation (WLI) with pseudo Nambu Goldstone bosons \cite{wliori1,wliori2,wliori3} and Minimal Warm Inflation (MWI) where the axions are coupled to non abelian gauge fields \cite{mwiori1,mwiori2,mwiori3}. The feasibility of the presently available models of WI has been acclaimed from the latest CMB estimates through the estimation of scalar spectral index and tensor to scalar ratio \cite{cqg2,long1,long2,long3,long4,long5}. The thermal effect from warm inflation has been found to lower the tensor to scalar ratio \cite{ori3, ori4, wliori1, long5, r1, r2, RDWI}. In this context, the effect of thermal and squeezing phenomena in WI is investigated with the recent joint analysis by $Planck$ and BICEP2/$Keck$. It can be shown that the combined thermal and squeezing effect in both strong and weak regimes of warm inflation can predict primordial thermal gravitational waves.
\\
\\
In warm inflation, the coefficient of dissipation is related to the temperature $(T)$  of the thermal bath as,
\begin{eqnarray}
\gamma(\phi,T) = A_{\gamma}T^d \phi^p M^{1-p-d}. \label{gam}
\end{eqnarray}
Here, $d$ and $p$ are integers that characterize the dependence of the dissipation coefficient on the temperature of the thermal bath and the inflaton field, respectively. Their values are determined by the microscopic interactions responsible for dissipative particle production and are therefore model dependent. $M$ is the cut-off scale of the particular warm inflationary model under consideration. The factor $M^{1-p-d}$ ensures that the dissipation coefficient has the appropriate mass dimension, in natural units. Consequently, allowed values and combinations of $d$ and $p$  are constrained by the underlying particle physics realization and thermal dynamics required for maintaining a warm inflationary regime. $A_\gamma$ is a dimensionless constant that carries the details of the microscopic model. The simplest potential that the inflaton ($\phi$) can take is $V(\phi)=\frac{1}{2}m^2\phi^2$. It is assumed that inflation proceeds in the slow roll regime and therefore, the first and second slow-roll parameters of the warm inflation can be expressed as
\begin{eqnarray}
\epsilon_H &=& \frac{-\dot H}{H^2}  = \frac{\epsilon_V}{1+Q}, \label{epwarm} \\ 
\eta_H &=& \frac{- \ddot \phi}{H \dot \phi}= \frac{1}{1+Q}(\eta_V - \beta + \frac{\beta -\eta_V}{1+Q}), \label{etwarm}
\end{eqnarray}
where, $\epsilon_V\simeq \frac{M_{pl}^2}{2}\bigg(\frac{V'(\phi)}{V(\phi)}\bigg)^2$,
$ \eta_V \simeq M_{pl}^2 \frac{V''(\phi)}{V(\phi)}$ and 
 $\beta = \frac{\gamma' V'(\phi)}{\gamma V(\phi)}$. The prime $(')$ and dot ($.$)  indicates derivative with respect to the field ($\phi$) and time ($t$) respectively. 
 One can see that WI implements important ingredients such as supersymmetry, string theory and involves the interaction of the inflaton with multiple fields. This interaction gives rise to thermal friction ($\gamma\dot\phi$) generated by the decay products, in addition to the Hubble friction due to expansion ($H\dot \phi$). This friction term allows slow-roll inflation to be sustained even for potentials with relatively larger curvature. As a result, the slow-roll conditions are relaxed, reducing the degree of fine-tuning required for the inflaton potential and alleviating the $\eta$-problem (detailed derivation of Eq.(\ref{gam}) through Eq.(\ref{etwarm}) can be found in Refs.\cite{ori4, RDWI, oyv, diss1, diss2}). The ratio of thermal and Hubble friction is represented by $Q = \frac{\gamma}{3H}$.   
The thermal fluctuations during the warm inflation modify the associated scalar power spectrum \cite{RDWI},
\begin{eqnarray}\label{ps}
P_{S_{d \neq 0}}^{WI} = \bigg(\frac{H^2}{2\pi \dot\phi} \bigg)^2 \bigg[\coth(\frac{H}{2T}) + \frac{T}{H} \frac{2\sqrt{3}\pi Q}{\sqrt{3+4\pi Q}}\bigg]  G_{d}(Q),
\end{eqnarray}
where,
\begin{eqnarray}\label{Gq}
G_{d}(Q)=\begin{cases}1+4.981Q^{1.946}+0.127Q^{4.336}, d=3\cr
1+0.335Q^{1.364}+0.0185Q^{2.315},  d=1\cr
\frac{1+ 0.4 Q^{0.77}}{(1+Q^{1.09})^2},  d=-1.\end{cases}
\end{eqnarray}
The various couplings of the inflaton to other fields can lead to an enhanced scalar power spectrum, thus leading to a decrease in the tensor to scalar ratio. 
 Therefore the corresponding scalar spectral index ($n_S^{WI}$) and the tensor to scalar ratio ($r^{WI}$) is given by \cite{cqg2, oyv, RR},
\begin{eqnarray}
1-n_S ^{WI} &=& \frac{1}{1+Q} \bigg\{4\epsilon_V - 2 \bigg( \eta_V - \beta + \frac{ \beta - \epsilon_V}{1+Q} \bigg) +  \frac{\alpha}{1+\alpha} \bigg[\frac{2\eta_V+\beta-7\epsilon_V}{4} +  \\&& \nonumber \frac{6+(3+4\pi)Q}{(1+Q)(3+4\pi Q)} \ (\beta-\epsilon_V)\bigg] \bigg\} - \frac{\mathrm{d}G_{d }(Q)}{\mathrm{d} \ \mathrm{ln}k},\\ 
r^{WI} &=&\frac{P_{T}^{WI}}{P_{S_{d \neq 0}}^{WI}} = \frac{16 \epsilon_V}{[\coth( \frac{H}{2T})+ \alpha]G_{d }(Q)}, \label{rwarm}
\end{eqnarray}
where, $\alpha =  \frac{T}{H}\frac{2\sqrt{3}\pi Q}{\sqrt{3+4\pi Q}}$. 
%and $\coth( \frac{H}{2T})=1+ 2n_k$. Here, $n_k$ is the distribution of inflaton particles in the thermal bath. For CI, $n_k =0$ and for WI it takes Bose-Einstein distribution \cite{RR},
 % \begin{eqnarray}
%n_k = \frac{1}{\exp(\frac{k}{ST})-1}.
% \end{eqnarray} 
In the strong dissipation regime ($Q>>1$ , $\alpha>>1$), the dynamics depart significantly from the standard CI. Similarly, the weak dissipation regime ($Q<<1$, $\alpha<<1$) leads to $G_{d }(Q)\simeq 1$ and $\frac{\mathrm{d}G_{d }(Q)}{\mathrm{d \ ln}k} = 0$.\\
\\
It is interesting to note that unlike CI, the contribution to the cosmological perturbations can come from radiation in the thermal bath \cite{sps1, sps2}. In WI, apart from the quantum fluctuations, the radiation contributes to thermal fluctuations that add to scalar perturbations (see Eq.(\ref{ps})).
Also, Eq.(\ref{rwarm}) is derived based on the assumption that metric perturbations in WI are negligible and the tensor power spectrum is retained as it is in the WI set up, i.e, $P_{T}^{CI} =  P_{T}^{WI}$ \cite{cqg2,sps1,sps2}. However, the principal objective of this work is to show that thermal gravitational waves can be sourced from WI. The gravitons are primarily produced from quantum fluctuations, however it may acquire a thermal effect during warm inflation. The thermal effect from the dissipation during WI on the tensor power spectrum and thermal gravitational waves remains unexplored so far with various observations.   The possibility of producing thermal gravitons from warm inflation is discussed in Ref. \cite{anx}. In that case, it is necessary to investigate the graviton production from the quantum fluctuations with thermal effect. In this work, the temperature dependent dissipation effect are naturally incorporated into the tensor power spectrum by placing the gravitons produced during inflation in a thermal squeezed vacuum. In this paper, for the first time, a connection between WI and thermal squeezing for TGWs is proposed and compared with the joint analysis data of $Planck$ and BICEP2/$Keck$ missions.  \\
 
 \section{Warm inflation and thermal squeezing}
In the present work, the gravitons produced from the squeezing phenomena are subjected to a thermal bath in a warm inflation and are placed in a thermal squeezed vacuum state for a quasi de Sitter expansion.
Consider a perturbed FLRW metric $ds^2 =- S^2(\tau)[d\tau^2-(\delta_{ij}+h_{ij})dx^i dx^j] $ where $ S(\tau)$ is the conformal time dependent scale factor and $h_{ij}$ is the metric perturbation. The tensor perturbation has two polarization states ($+$ and $\times $) denoted by the superscript $z$ which can be represented in the operator form as,
\begin{eqnarray}\label{hjk}
h^{(z)}(x,\tau)&=&\frac{\sqrt{16\pi}}{S(\tau)M_{pl}}\int \frac{d^3k}{(2\pi)^{3/2}}[a_kf_k(\tau)+  a_{k'}^{\dagger} f_k^*(\tau)]e^{ik.x}, \\
&=& \int \frac{d^3k}{(2\pi)^{3/2}} h_k(\tau)e^{ik.x},
\end{eqnarray}
where $a_k, a_{k'}^{\dagger}$ are the annihilation and creation operators respectively and $f_k(\tau), f_k^*(\tau)$ are the mode functions. The evolution of  $f_k(\tau)$ is given by  the Mukhanov-Sasaki equation, \begin{equation}  f_k(\tau)''+ (k^2 - \frac{S''}{S}) f_k(\tau) =0. \end{equation}
In a quasi de Sitter universe, during inflation, the scale factor and the conformal time are related by $S(\tau) = \frac{-1}{H\tau(1- \epsilon)}$, therefore the mode function obeys,  \begin{equation} f_k(\tau)''+ [k^2 - \frac{1}{\tau^2}(\nu^2-\frac{1}{4})] f_k(\tau) =0, \end{equation} where, $\nu=\frac{3}{2}+\epsilon$ for small values of $\epsilon$.
The general solution of this equation is obtained in terms of the Hankel functions $H_\nu^{(1)}$ and $H_\nu^{(2)}$ as, \begin{equation} f_k(\tau)=\sqrt{-\tau}[c_1(k) H_\nu^{(1)}(-k\tau) + c_2(k) H_\nu^{(2)}(-k\tau)]. \label{suph} \end{equation}
For the subhorizon modes ($k>>SH$), $f_k(\tau)=\frac{1}{\sqrt{2k}}e^{-ik\tau}$ and for the superhorizon modes ($k<<SH$) \cite{Riotto},
\begin{equation}
f_k(\tau)=e^{i(\nu-1/2)(\pi/2)}2^{\nu-3/2}\frac{\Gamma(\nu)}{\Gamma(3/2)}\frac{1}{\sqrt{2k}}(-k\tau)^{1/2 - \nu}. \label{mse}
\end{equation} In Eq. (\ref{suph}), $-k\tau>>1$ implies superhorizon modes.
Usually, the tensor power spectrum ($P_T^{CI}$) is computed based on the assumption that the gravitons occupy the zero temperature vacuum state $\ket{0}$,
\begin{eqnarray} 
\braket{0|h_kh_{k'}|0}&=& \frac{2\pi^2}{k^3}P_T^{CI}\delta^3(k-k').
\end{eqnarray}
The gravitons produced during WI are assumed to remain in the thermal bath rather than to settle in an initial zero temperature vacuum. Any temperature dependent inflationary model must be evaluated in a thermal state rather than in an ordinary vacuum. A thermal operator
\begin{eqnarray}
\mathcal{T}(\theta_k)&=&e^{-\theta_k(a_k \tilde a_k -a_k^{\dagger} \tilde a_k^{\dagger})},
\end{eqnarray}
can be used to create such a state. Since the gravitons are produced by quantum fluctuations, they can be inherently associated with a squeezed state description  \cite{ss1, ss2, ss3, ss4, ss5, ss6, pre1, pre2, pre3}.
A squeezing operator in quantum optics is defined as,
\begin{eqnarray}
\mathcal{Z}(q,\zeta)&=&e^{\frac{1}{2} [q(e^{-2i\zeta}a_ka_k-e^{2i\zeta} a_k^\dagger a_k^\dagger)]}.
\end{eqnarray}
The inflaton interactions in WI can produce pairs of particles that naturally form a double mode squeezed state \cite{dsv1, dsv2, dsv3, dsv4}. Therefore, in a thermal bath containing radiation, it is essential to modify the conventional inflationary vacuum using suitable Bogoliubov transformations \cite{gris}. A thermal squeezed vacuum state ($\ket{\mathcal{T}_{SV} }$)  is defined in terms of a quantity representing the average number of particles ($\theta_k$), squeezing parameter ($q$) and squeezing angle ($\zeta$) as,
\begin{eqnarray}
\ket{\mathcal{T}_{SV} }&=& \mathcal{T}(\theta_k)\mathcal{Z}(q,\zeta) \mathcal{\tilde Z}(\tilde q, \tilde \zeta)\ket{0 \ \tilde 0}.\label{tsv}
\end{eqnarray}
Here, all the operators with a tilde (for eg. $ \tilde \zeta, \tilde a,..$etc.) are their corresponding counterparts in the extended Hilbert space according to the thermofield dynamics \cite{book1,book2,book3,book4,book5,book6}. \\ \\
This paper focuses on the possibility of producing gravitons with thermal effect and hence the corresponding TGWs during warm inflation. Therefore, a thermal squeezed vacuum can be used to find the tensor perturbations using Eqs. (\ref{hjk}), (\ref{mse}) and (\ref{tsv})
\begin{eqnarray}
\braket{\mathcal{T}_{SV}|h_kh_{k'}|\mathcal{T}_{SV}} = \frac{16\pi}{S^2 (\tau) M_{pl}^2}  [(1+2 \sinh^2q)|f_k|^2  + \frac{1}{2}\sinh2q (e^{i\zeta}f_k^2+e^{-i\zeta}f_{k}^{*2})]  \coth(\frac{k}{2T}) \delta^3(k-k'), \label{18}
\end{eqnarray}
and thus the tensor power spectrum is
\begin{eqnarray}
P^{TSV}_T(k) &=& \frac{16\pi}{M_{pl}^2} \bigg(\frac{H}{2\pi} \bigg)^2\bigg(\frac{k}{SH} \bigg)^{n_T}  \bigg[1+2 \sinh^2q +  \sinh2q  \cos(\zeta+(\nu-\frac{1}{2})\pi)\bigg]  \coth(\frac{k}{2T})\\
&=& A_T(k_0) \bigg(\frac{k}{k_0} \bigg)^{n_T}  \bigg[1+2 \sinh^2q +  \sinh2q  \cos(\zeta+(2-n_T)\frac{\pi}{2})\bigg] \coth(\frac{k}{2T})\\ \label{ptsv}
&=&P_T^{CI}(k) \bigg[1+2 \sinh^2q +  \sinh2q \cos(\zeta+ (2-n_T) \frac{\pi}{2})\bigg]  \coth(\frac{k}{2T}).
\end{eqnarray}
Here, $1+2 \sinh^2q + \sinh2q \cos(\zeta+(2-n_T)\frac{\pi}{2})$ is the squeezing factor ($SF$), $\coth(\frac{k}{2T})$ is the factor that accounts for thermal contribution from WI. The thermal squeezed spectrum is not the only contribution to the tensor spectrum of warm inflation. In warm inflation, the presence of a radiation bath can influence tensor perturbations through both the modified background dynamics and the anisotropic stress of the thermalized matter sector. In the present analysis, the latter contribution is neglected by assuming that the transverse traceless anisotropic stress is sufficiently small, so that the tensor modes evolve as free gravitational perturbations on the warm inflationary background. The effect of dissipation from warm inflation is therefore incorporated through the background quantities, particularly the Hubble parameter and the radiation energy density through the dissipation ratio $Q=\gamma/3H$. Subsequently, a thermal squeezed initial state is considered for the tensor modes. It is important to emphasize that the thermal squeezed state represents an assumed initial state modification and should not be interpreted as a complete description of all thermal effects on tensor perturbations. Under these assumptions, the primordial tensor spectrum is obtained by combining the warm inflationary background evolution ($Q$ factor) with the thermal and squeezing factors associated with the initial state. \\
\\
In warm inflation, the modes freeze out when their physical wave number becomes smaller than the freeze out scale ($k \leq k_F$). The wave number at freeze out is given by 
$ k_F =\sqrt{H\gamma}$ for $Q>>1$ and $k_F \sim H$ for $Q<<1$. The thermal friction $\gamma$ can be expressed in terms of $H$ and $Q$ through the relation $\gamma = 3HQ$. Substituting, for $k=k_F$ and $\gamma$ in $\coth$ term,
\begin{equation}
\coth(\frac{k}{2T}) =\begin{cases} \coth ( \frac{\sqrt{3Q}H}{2T} ), Q>>1,\\
\coth ( \frac{H}{2T} ), Q<<1. \end{cases}\label{cot}
\end{equation}
Since the gravitons produced during inflation decouple earlier than radiation, their temperature may be slightly reduced. But this may not significantly
affect the overall temperature of the thermal bath as the effective number of relativistic degrees of freedom of
the particles ($g_*$) involved is large \cite{thesis}. The corresponding energy density of the thermal bath attains a steady state during WI, \begin{equation} \rho_r = \frac{\pi}{30}g_* T^4 = \frac{3}{4}Q\dot\phi^2. \label{re} \end{equation}
Using Eq. (\ref{re}) in the Friedmann equation ($H^2 \propto \rho$) and relating it with the equation of motion of inflaton in a slow roll setting $3H(1+Q)\dot\phi = V'(\phi)$, $\frac{T}{H}$ can be expressed in terms of $Q$  \cite{cqg2} as,
\begin{equation}\label{tH}
\frac{T}{H} =   \bigg( \frac{135}{64} \frac{\epsilon_V}{\pi^4 V(\phi) g_*} \bigg)^{\frac{1}{4}} \bigg[ \frac{Q}{(1+Q)^2} \bigg]^{\frac{1}{4}}.
\end{equation}
The Eqs. (\ref{cot}) and (\ref{tH}) can be incorporated in Eq. (\ref{ptsv}) and the tensor power spectrum in thermal squeezed vacuum for WI becomes,
\begin{eqnarray} \label{ptsv2}
P^{TSV}_T(k) = P_T^{CI}(k)\times SF \times \begin{cases}
         \coth\bigg[\frac{\sqrt{3Q}}{2c} \bigg(\frac{(1+Q)^2}{Q}\bigg)^{\frac{1}{4}}\bigg], \text{for } Q \gg 1\cr
         \coth\bigg[\frac{1}{2c} \bigg(\frac{(1+Q)^2}{Q}\bigg)^{\frac{1}{4}}\bigg], \text{for } Q \ll 1\end{cases}
\end{eqnarray}
where, $c= \bigg( \frac{135}{64} \frac{\epsilon_V}{\pi^4 V(\phi) g_*} \bigg)^{\frac{1}{4}}$.\\
\\
 The occupation number of gravitons in warm inflationary scenario can be determined from the kinetic Boltzmann equation of the form
\begin{eqnarray}
\frac{\partial n_k(t)}{\partial t}
-Hk\frac{\partial n_k(t)}{\partial k}
=C_g[n_k].
\end{eqnarray}
Here $n_k(t)$ denotes the graviton phase-space distribution and $C_g[n_k]$ contains the relevant production, absorption and scattering processes in the thermal bath.
The collision term can be expressed as
\begin{eqnarray}
C_g[n_k]= -\Gamma_g(k,T)[n_k(t)-n_k^{eq}]
\end{eqnarray}
where $\Gamma_g$ is the graviton interaction rate and the equilibrium distribution of gravitons can be described using Bose-Einstein statistics
\begin{eqnarray}
n_k^{eq}=\frac{1}{e^{\frac{k}{T}}-1}.
\end{eqnarray}
\\
\noindent The Boltzmann equation can be rewritten interms of $\Gamma_g$ as,
\begin{eqnarray}
\frac{\partial n_k(t)}{\partial t} -Hk \frac{\partial n_k(t)}{\partial k} = -\Gamma_g(k,T)[n_k(t)-n_k^{eq}].
\end{eqnarray}
The inverse of $\Gamma_g$ is roughly the mean time between effective interactions.
Therefore, for large $\Gamma_g$, the gravitons interact frequently and for small $\Gamma_g$, the gravitons interact rarely.
The relevant comparison in an expanding universe is $\frac{\Gamma_g}{H}$.
For efficient thermalization, $\Gamma_g/H \gg 1$ whereas $\Gamma_g/H\ll1$ corresponds to a decoupled graviton population. This distinction is important because the radiation bath can remain thermalized even when the gravitons are not. A thermal distribution can be regarded as a consistent limiting solution only when the corresponding interaction rate is sufficiently large compared with the expansion rate, or when the source and relaxation terms dynamically drive the system toward such a distribution, i.e., $n_k \rightarrow n_k^{eq}$.
The characteristic expansion timescale is $t_H\sim\frac{1}{H}$, while the graviton interaction timescale is $t_{\rm int}\sim\frac{1}{\Gamma_g}$.
Therefore, $\Gamma_g\gg H$
implies $t_{\rm int}\ll t_H$.
In this regime, gravitons undergo many interactions within one Hubble time and can approach thermal equilibrium ($n_k\simeq n_k^{\rm eq}$).
Conversely, $\Gamma_g\ll H$ means that cosmic expansion is faster than graviton interactions. The gravitons then effectively decouple from the thermal bath, and a thermal distribution cannot be assumed without additional justification.\\
\\
The two-point correlation function of tensor perturbations already contains the thermal occupation number,
\begin{eqnarray}
\braket{\mathcal{T}_{SV}|h_kh_{k'}|\mathcal{T}_{SV}} &\propto& \coth\frac{k}{2T} \times SF \nonumber \\
&\propto& [2 \frac{1}{e^{\frac{k}{T}}-1} +1] \times SF \nonumber  \\
&\propto& \left[2n_k^{eq}+1\right] \times SF \label{be}
\end{eqnarray}
This shows that the thermal factor in Eq. (\ref{18}) and hence in Eq. (\ref{cot}) is related to the kinetic evolution of $n_k$.
Therefore, thermal squeezing and dissipation in warm inflation allows us to distinguish the tensor modes that generate the primordial gravitational waves with thermal features, which are a remnant of WI. The standard CI can be recovered in the limit $Q\rightarrow 0$ and $q\rightarrow0$, allowing a strong coherence between the microphysics of WI and CI using the corresponding mathematical formulations. This modification to the tensor modes from the WI must reflect on the BB mode angular power spectrum of CMB and may shed some light on the primordial thermal gravitational waves generated from warm inflation. 

\section{BB mode angular power spectrum}
\begin{figure}[h!]
\centering
\makebox[\linewidth]
{
   \begin{subfigure}{0.45\linewidth}
    \includegraphics[width=\linewidth]{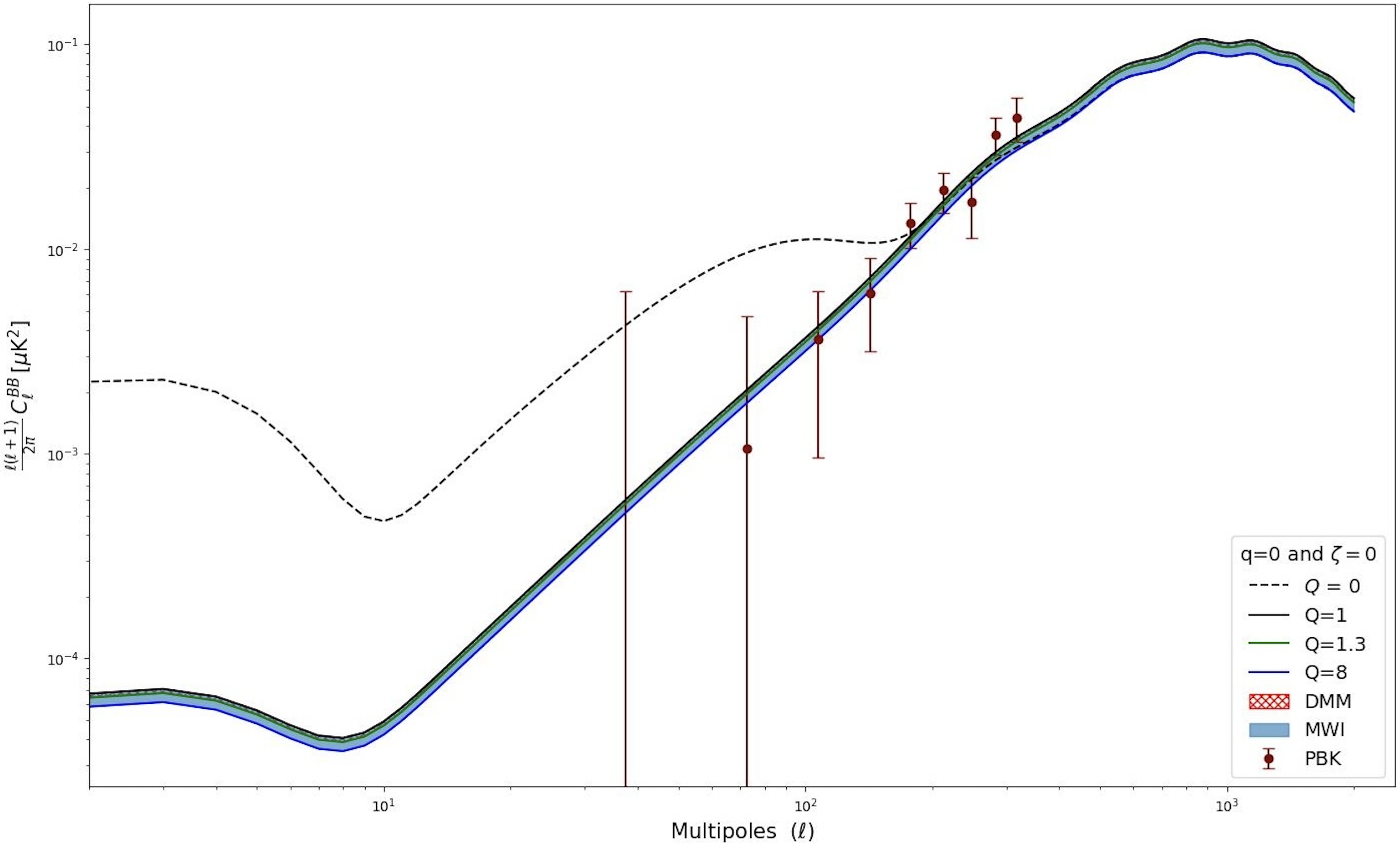}
     \caption{$q=0$ and $\zeta=0$}
     \end{subfigure}
     \begin{subfigure}{0.45\linewidth}
    \includegraphics[width=\linewidth]{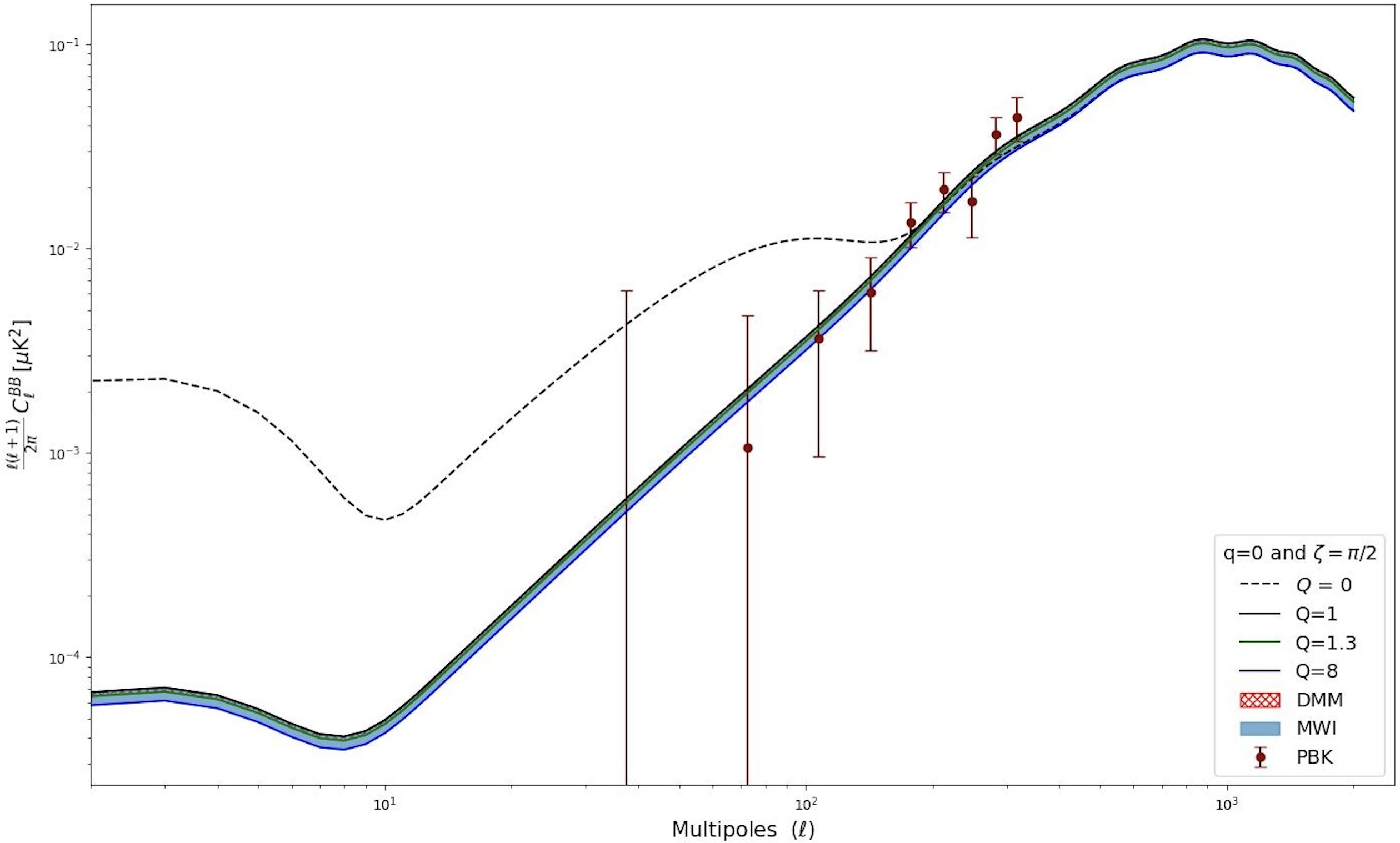}
    \caption{$q=0$ and $\zeta=\pi/2$}
     \end{subfigure}
}\par
\makebox[\linewidth]
{
   \begin{subfigure}{0.45\linewidth}
    \includegraphics[width=\linewidth]{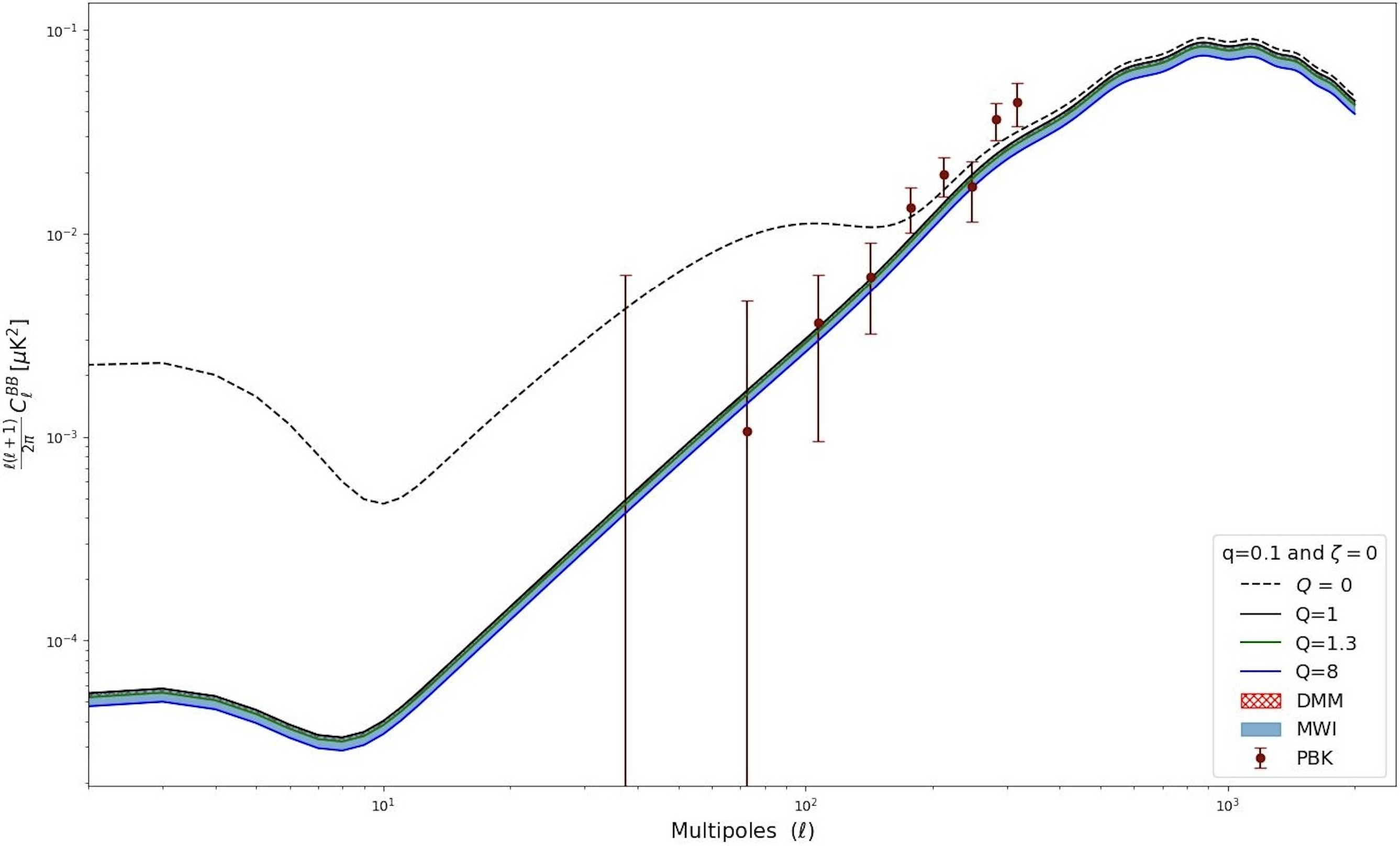}
     \caption{$q=0.1$ and $\zeta=0$}
     \end{subfigure}
     \begin{subfigure}{0.45\linewidth}
    \includegraphics[width=\linewidth]{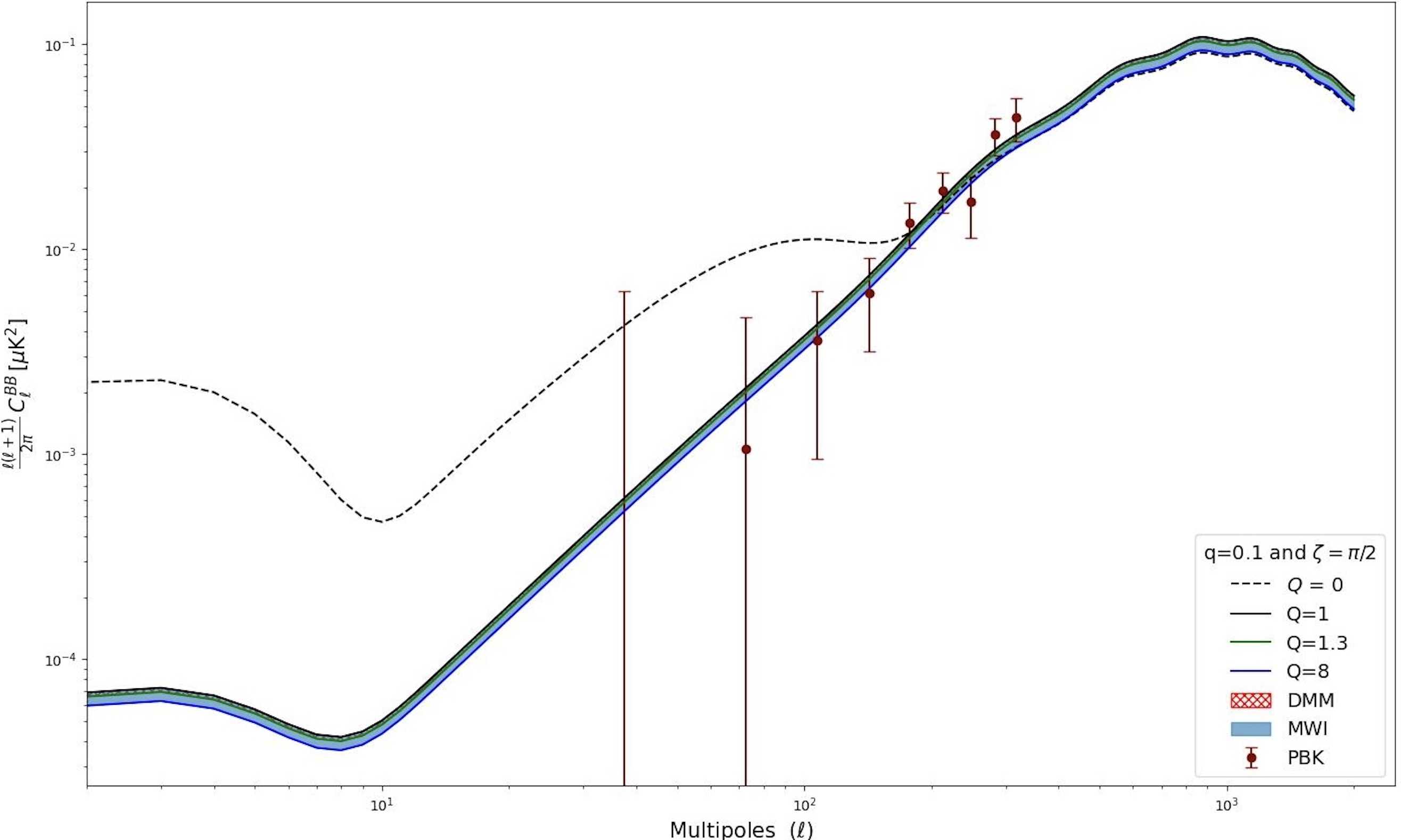}
    \caption{$q=0.1$ and $\zeta=\pi/2$}
     \end{subfigure}
}
\makebox[\linewidth]
{
   \begin{subfigure}{0.45\linewidth}
    \includegraphics[width=\linewidth]{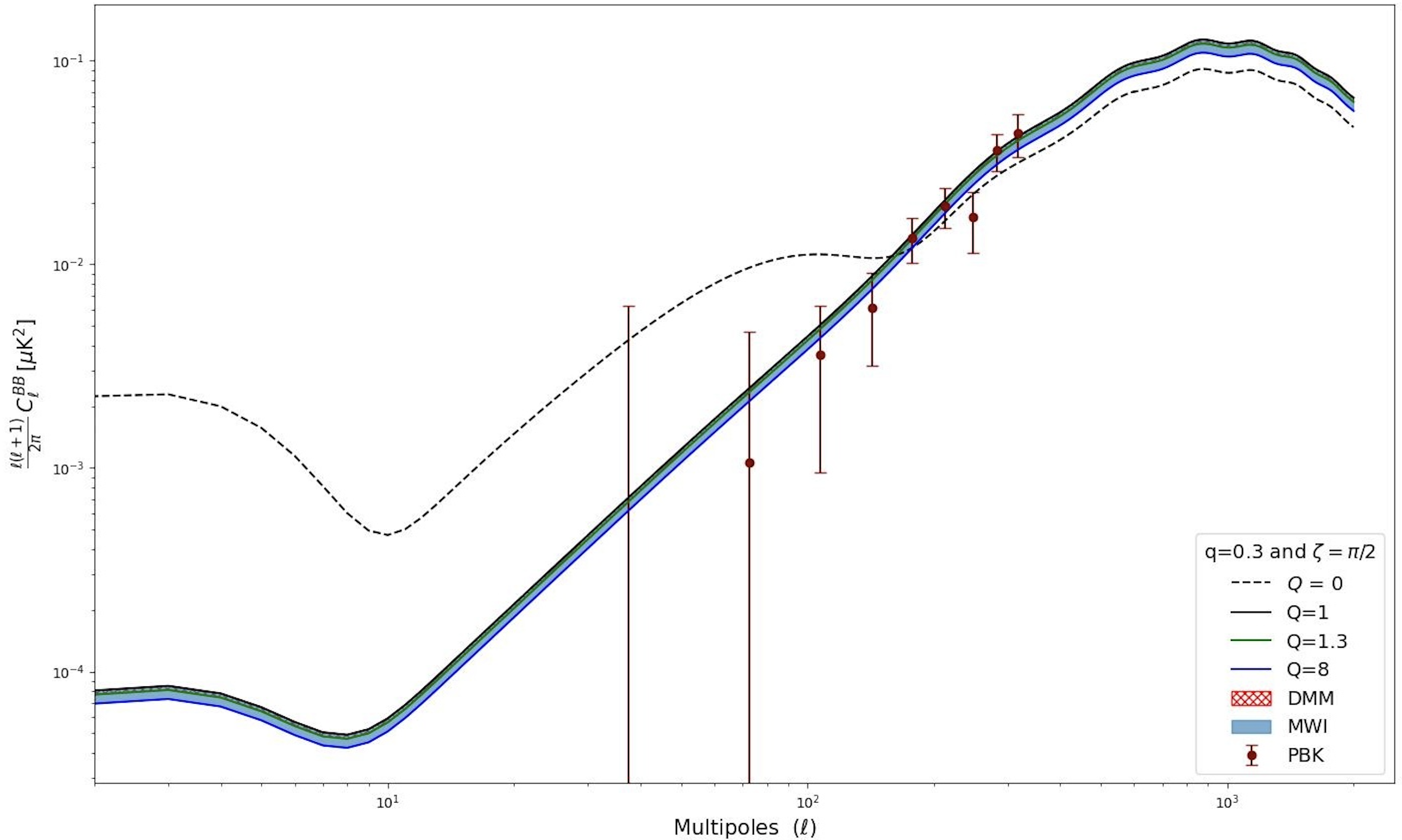}
     \caption{$q=0.3$ and $\zeta=0$}
     \end{subfigure}
     }
\caption{\label{strong} BB mode correlation angular power spectrum for strong dissipative ($Q>1$) warm inflation models in thermal squeezed vacuum state for various squeezing strength and angle with joint analysis of $Planck$ and BK18 data.  }
\end{figure}

\begin{figure}[h!]
\centering
\makebox[\linewidth]
{
   \begin{subfigure}{0.45\linewidth}
    \includegraphics[width=\linewidth]{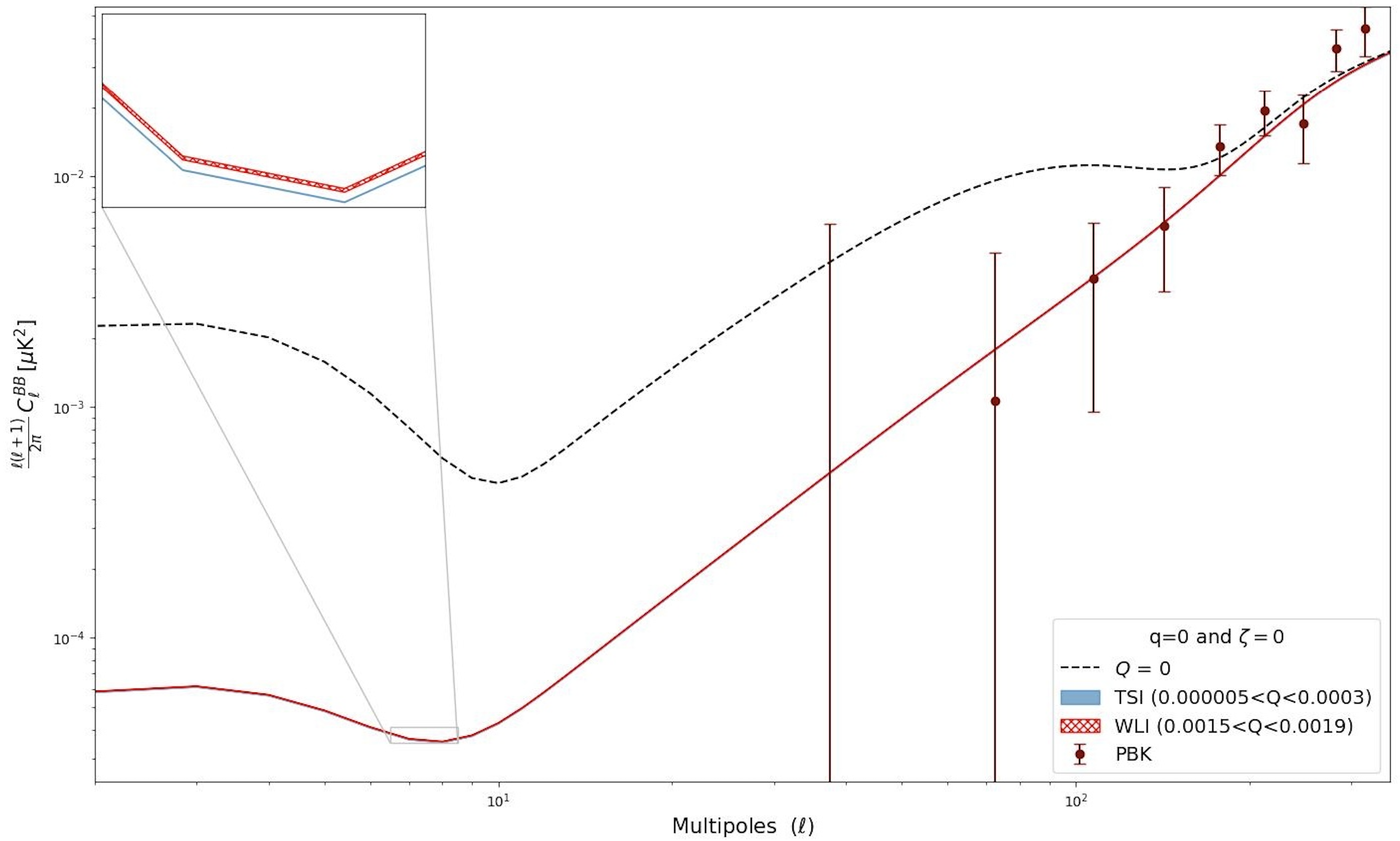}
     \caption{$q=0$ and $\zeta=0$}
     \end{subfigure}
     \begin{subfigure}{0.45\linewidth}
    \includegraphics[width=\linewidth]{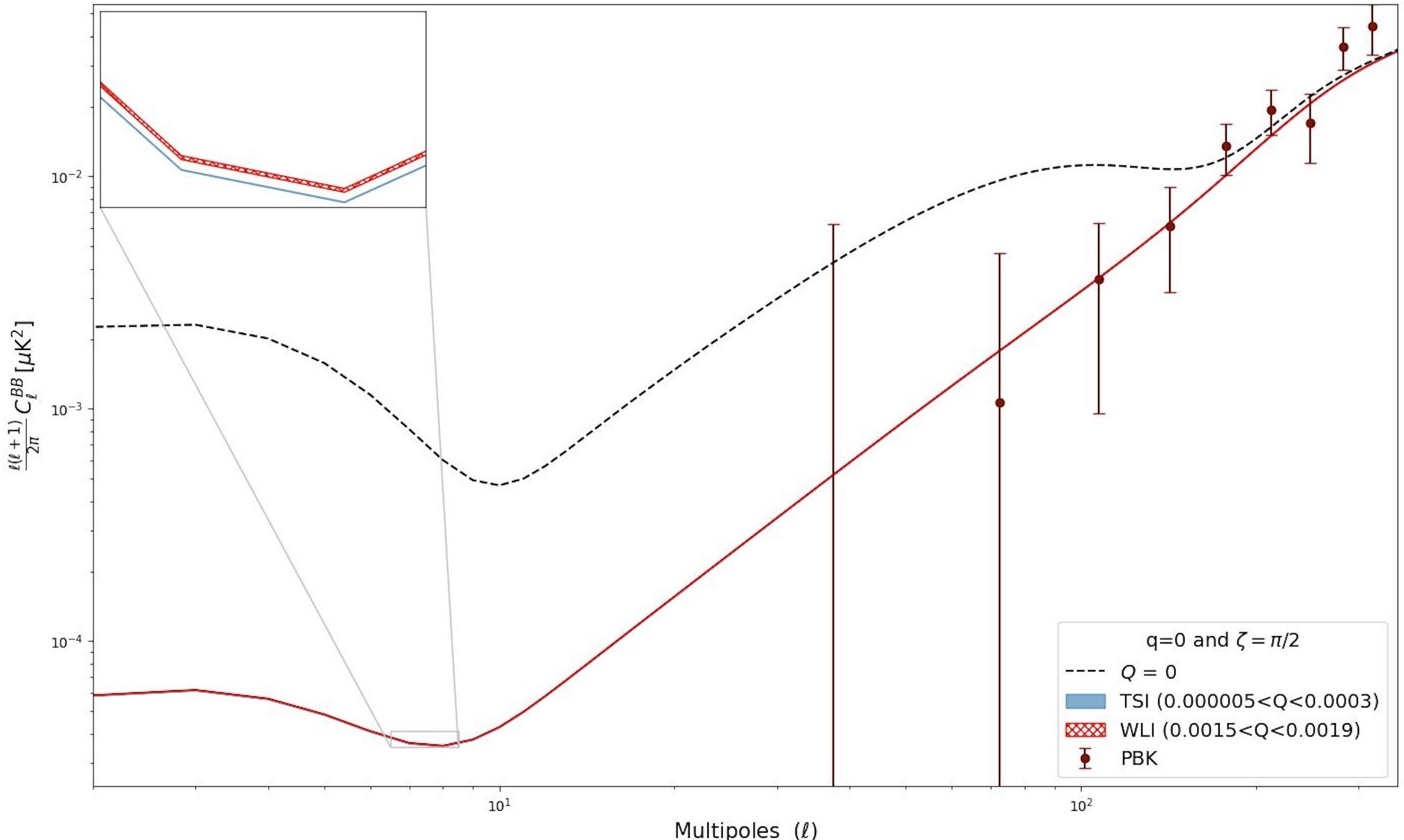}
    \caption{$q=0$ and $\zeta=\pi/2$}
     \end{subfigure}
}\par
\makebox[\linewidth]
{
   \begin{subfigure}{0.45\linewidth}
    \includegraphics[width=\linewidth]{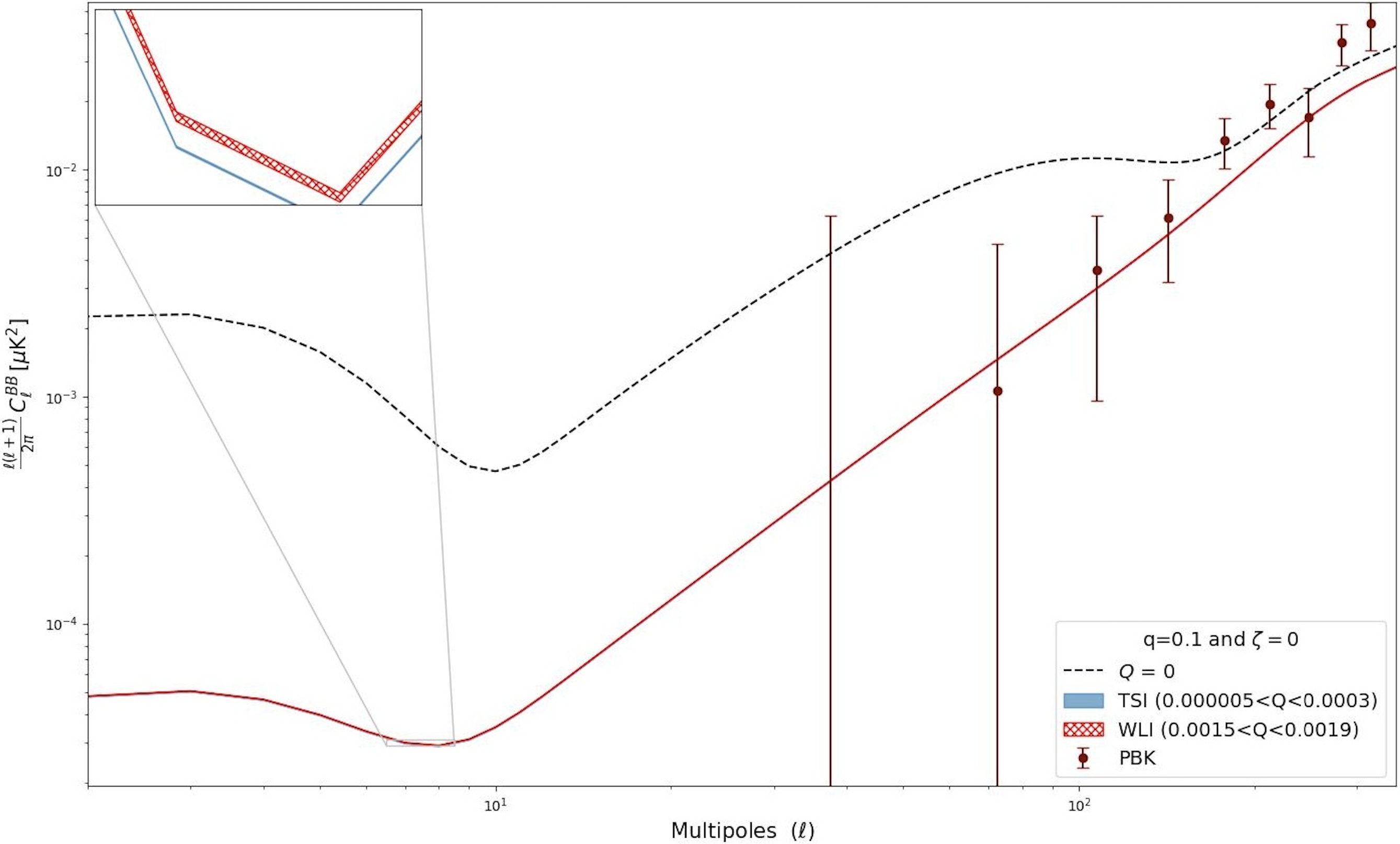}
     \caption{$q=0.1$ and $\zeta=0$}
     \end{subfigure}
     \begin{subfigure}{0.45\linewidth}
    \includegraphics[width=\linewidth]{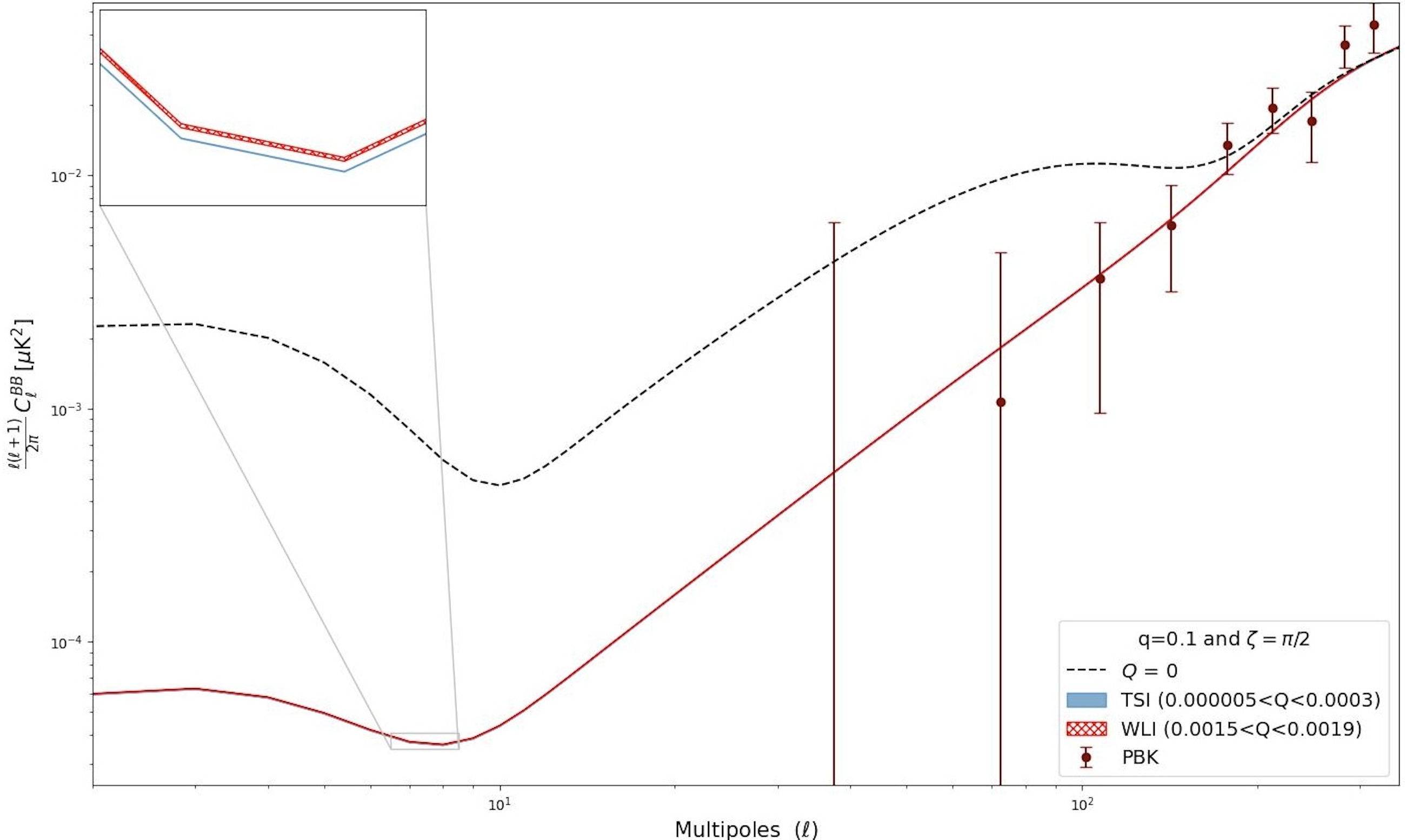}
    \caption{$q=0.1$ and $\zeta=\pi/2$}
     \end{subfigure}
}
\makebox[\linewidth]
{
   \begin{subfigure}{0.45\linewidth}
    \includegraphics[width=\linewidth]{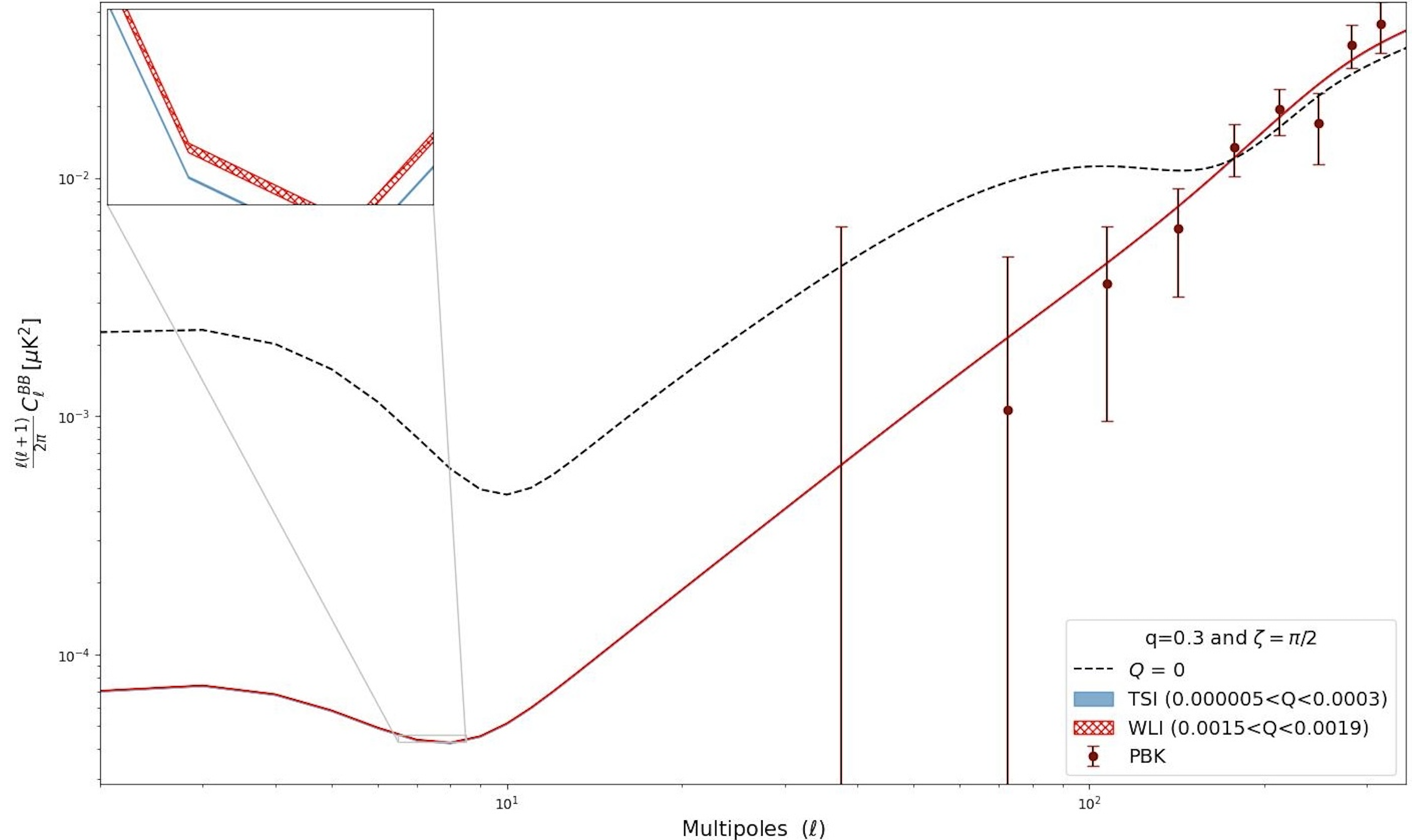}
     \caption{$q=0.3$ and $\zeta=\pi/2$}
     \end{subfigure}
}
\caption{\label{weak} BB mode correlation angular power spectrum for weak dissipative ($Q<1$) warm inflation models in thermal squeezed vacuum state for various squeezing strength and angle with joint analysis of $Planck$ and BK18 data.  }
\end{figure}

One of the major difficulties in the search for B-modes is the contribution from galactic foregrounds that mimics the primordial signals. The dust emission can be distinguished from the inflationary signals by a proper choice of dust correction model. The dust B-modes are separated from the inflationary B-modes through the different dust frequency dependence of $Planck$ (detected at 353 GHz) and BICEP2/$Keck$  (detected at 150GHz) by adopting a similar approach followed in the joint analysis of $Planck$ and BICEP2/$Keck$  \cite{Ade}. The auto and cross-spectra $\frac{BK \times BK - \mu BK \times P}{1-\mu}$, conveniently denoted by PBK, are evaluated at a fiducial value of the model parameter $\mu=0.04$, which may clean out the dust contribution from BK18\footnote{Here, BK18 refers to the data collected by BICEP2/$Keck$  in the 2018 observational season and released in 2021.} data  \cite{BK18}. The modified primordial tensor power spectrum in Eq. (\ref{ptsv2}) can be used to compute the BB mode angular power spectrum in a warm inflationary setting using the following relation,
\begin{eqnarray}
\frac{C_{l}^{BB}}{4\pi^2} =& \int dk \ k^2 \ P^{TSV}_T(k) \times \bigg|\int_0^{\tau_0} d\tau g(\tau) f_{k}(\tau)  \times  \bigg[2j_l'(x) + \frac{4j_l(x)}{x} \bigg] \bigg|^2. \label{bb}
\end{eqnarray}
Here, the visibility function $g(\tau) = \kappa(\tau)' e^{-\kappa(\tau)}$ is expressed in terms of optical depth ($\kappa$) and its differential ($\kappa'$). $\tau$ is the conformal time. Hence, analysing Eqs. (\ref{18}-\ref{be}), it can be shown that the thermal graviton occupation number for warm inflation enters Eq. (\ref{bb}) through Eq. (\ref{ptsv2}). For a quadratic inflation with e-foldings $N=60$, the DMM model is found to be consistent with the joint analysis of $Planck$ and BICEP2/$Keck$ for $1 < Q < 1.3$ and MWI for $1<Q<8$ \cite{cqg2}. Similarly, the scalar spectral index and tensor to scalar ratio of quadratic inflation for TSI and WLI fall within the $68\%$ CL of joint $Planck$-BK18 data for $5 \times 10^{-6} < Q < 0.0003$ and $0.0015 < Q < 0.0019$ respectively \cite{cqg2}. Therefore, in the current study, DMM and MWI are taken as the working model for WI in the strong dissipative regime ($Q\gg1$), whereas TSI and WLI represent WI in the weak dissipative regime ($Q\ll1$). In the present work, DMM \cite{ori1, dmmori1, dmmori2, DMMnew1}, MWI \cite{mwiori1, mwiori2, mwiori3}, TSI \cite{tsiori1, tsiori2} and WLI \cite{wliori1, wliori2, wliori3} with the following interactions are considered
\begin{eqnarray}
\mathcal{L}_{DMM} &=& \sum_{j=1}^{N_M} \sum_{k=1}^{N_\chi} { \frac{1}{2} (\partial_\mu \chi_{jk})^2- \frac{b_{jk}}{4!} \chi_{jk}^{4} -  \frac{y_{jk}^{2}}{2} ( \phi - M_{j})^2 \chi_{jk}^{2} } +  \sum_{j=1}^{N_M} \sum_{k=1}^{\frac{N_\chi}{4}} { i \bar \psi_{jk} \not{\partial} \psi_{jk} - z_{jk}(\phi - M_j) \bar \psi_{jk} \psi_{jk}}, \nonumber\\ 
\mathcal{L}_{MWI} &=& \frac{u}{16 \pi} \frac{\Phi}{\lambda} \tilde G^{\mu \nu}_i G^i_{\mu \nu}, \nonumber\\
\mathcal{L}_{TSI} &=& b^2 |\chi|^4 +z^2 |\sigma|^4 +   4b^2 |\chi|^2 |\phi|^2  +  4bz Re[\phi^{\dagger} \chi^{\dagger} \sigma^2] +  4z^2 |\chi|^2 |\sigma|^2, \nonumber\\
\mathcal{L}_{WLI} &=& \frac{g}{\sqrt{2}} (-\phi_1 - \phi_2)\bar \psi_{1L}\psi_{1R} - \frac{ig}{\sqrt{2}} (\phi_1 - \phi_2)\bar \psi_{2L}\psi_{2R} - h\sigma \sum_{i=1,2} (\bar \psi_{iL} \psi_{\sigma R} + \bar \psi_{\sigma L} \psi_{i R}),\nonumber
\end{eqnarray}
that yield the dissipation coefficients $\gamma_{DMM} \propto  \frac{\phi^2}{T}$,  $\gamma_{MWI} \propto \frac{u^5T^3}{\lambda^2}$, $\gamma_{TSI} \propto \frac{T^3}{\phi^2}$ and $\gamma_{WLI} \propto T $ respectively. Here, $b_{jk}$ = $y_{jk}$ = $z_{jk}$, $b$ and $u$ are the various coupling constants, $\lambda$ is the axion decay constant and $\mathcal{L}_{TSI}$  is the scalar interaction corresponding to the superpotential $W = b \Phi X^2 + z XY^2$.\\
\\ 
The Code for Anisotropies in the Microwave Background (\texttt{CAMB}) \cite{CAMB} was used to propagate the modified primordial tensor spectrum through the standard CMB transfer calculation. The thermal squeezed vacuum effects were incorporated at the level of the primordial tensor spectrum rather than modifications to the \texttt{CAMB} Boltzmann hierarchy. Using \texttt{CAMB}, the BB mode angular power spectrum in a thermal squeezed vacuum state is studied for various values of $Q$, the squeezing parameter range $0 \leq q \leq 1$ and $\zeta = 0,\pi/2$ in the strong and weak regimes with  $\kappa$ = 0.056 \cite{planck} in Eq. (\ref{ptsv2}). Here, the thermal effect is accounted through the  $Q$ dependent $\coth$ term. Comparison of the obtained results for various strong and weak regimes of warm inflationary models with dust corrected CMB observations is presented in Fig. (\ref{strong}) and (\ref{weak}), respectively.
\noindent A clear distinction between WI and CI can be observed on the amplitude of BB mode angular power spectrum of CMB for both weak and strong dissipation regimes. The BB mode angular power spectrum of CMB for WI has a lower amplitude compared to CI at lower multipoles, especially for $\zeta=0$ (see the subfigures (a) and (c) of Fig \ref{strong} and \ref{weak}). For the squeezing angle $\zeta=\pi/2$, the amplitude is found to gradually rise above that of CI at higher multipoles (see the subfigures (b), (d) and (e) of Fig \ref{strong} and \ref{weak}). It can be inferred from the analysis that the lowering of the amplitude is present even in the absence of the squeezing effect ($q=0$ and $\zeta=0$) (see Fig \ref{strong}(a) and \ref{weak}(a)). This implies that the thermal effect ($Q$) from WI can be solely responsible for the lowering of the amplitude of the BB mode angular power spectrum at lower multipoles. However, the squeezing effect ($q \neq 0$ or $\zeta \neq 0$), can bring the amplitude of the BB mode angular power spectrum of CMB for various warm inflationary models within the observational limits, especially at higher multipoles, suggesting the existence of TGWs in a warm inflationary scenario. The BB mode angular power spectrum of CMB for WI in a thermal squeezed state lies well within the bounds of dust corrected joint analysis of $Planck$ and BK18 in both strong and weak dissipation (see Fig \ref{strong}(d) and \ref{weak}(e)). Further scrutiny shows that neither the thermal effect nor the squeezing effect can alone account for the observational data. For WI in the strong dissipation regime ($Q>>1$), the thermal and squeezing effects $q=0.1$ and $\zeta=\pi/2$ are more compatible with the observations, whereas in the weak dissipation regime ($Q<<1$), $q=0.3$ and $\zeta=\pi/2$ are preferred by the observations. Therefore, it may be concluded that the combined thermal and squeezing effect in warm inflation makes it compatible with the recent CMB BB-mode observations. The present work does not perform any statistical parameter estimation or Bayesian model comparison. The analysis is based on comparing theoretical results with the recent observational constraints to assess the phenomenological viability of the proposed model.\\

\section{Conclusions}
A complete description of inflation and its successive stages requires an alternate scenario that can explain the kinematics involved and naturally accommodate the observable signatures. This can be achieved by introducing some extra effects in the conventional CI that can potentially alter the scenario.  The slow roll WI is found to be inherently capable of explaining the physics during inflation and subsequent events, such as particle production and reheating, within a multifield framework. Literature in WI so far claims that the tensor modes are not affected by the thermal bath and hence the tensor power spectrum retains the same form in WI as in CI. For the first time, the effect of thermal squeezed vacuum from WI on the B-modes is investigated. The inflationary frameworks for producing sufficient squeezing are explored and validated using joint $Planck$ and BK18 observations. The reduction in the primordial tensor power spectrum together with the enhancement of the scalar power spectrum can suppress the tensor to scalar ratio further, which manifests in the BB-mode. This decline in the amplitude of the BB-mode angular power spectrum of CMB due to the contribution from the thermal bath of the warm inflation suggests that primordial gravitational waves may be thermal in nature and therefore can be considered as the signal of thermal gravitational waves. The standard power spectrum in CI can be recovered in the absence of thermal effect ($Q=0$) and squeezing ($q=0$ and $\zeta=0$). Even though the results are consistent with observations for both strong and weak dissipative models, MWI may be preferred over the other models for its wider range of BB mode angular power spectrum that may accommodate more future observations. Thus, the joint analysis of 
$Planck$ and BK18 does not rule out the multifield warm inflationary scenario. The analysis of the BB mode angular power spectrum of the selected warm inflationary models provides a satisfactory fit to the current observational data. However, a comprehensive statistical comparison with alternative inflationary scenarios using Bayesian evidence or information criteria is beyond the scope of the present work.
The MWI assumes axionic coupling and may open a window to search for a dark matter candidate. Moreover, superstring theory based WI models involve multiple fields, any one of which can be a dark matter candidate. Thus, the experimental detection of any of the aforementioned effects would also help in bridging cosmology and particle physics. These promising qualities, along with the key result of this article, can bring WI into the limelight for probing the early universe with CMB.

\section*{Acknowledgments}
AB acknowledges the financial support of Prime Minister’s Research Fellowship (PMRF ID : 3702550) provided by the Ministry of Education, Government of India. AB is grateful to Prof. P K Suresh, School of Physics, University of Hyderabad, for constant support, motivation, useful discussions and valuable suggestions.

\end{document}